\documentclass[aps,prl,reprint,groupedaddress]{revtex4-1}

\usepackage{graphicx, amsmath, epstopdf, soul, color}

\begin{document}



\title{Counter-propagating radiative shock experiments on the Orion laser}


\author{F.~Suzuki-Vidal}
\email[]{f.suzuki@imperial.ac.uk}

\author{T.~Clayson}
\author{G. F.~Swadling}\altaffiliation{Currently at Lawrence Livermore National Laboratory, CA, USA}

\author{S. V. Lebedev}
\author{G. C. Burdiak}

\affiliation{Blackett Laboratory, Imperial College London, SW7 2BW, UK}

\author{C.~Stehl\'e}
\author{U.~Chaulagain}\altaffiliation{Currently at ELI, IOP-CAS, Prague, Czech Republic.}
\author{R.L. Singh}
\affiliation{LERMA UMR 8112,  Observatoire de Paris, PSL Research University, UPMC , Sorbonne University, CNRS, France}

\author{J. M.~Foster}
\author{J.~Skidmore}\altaffiliation{Currently at First Light Fusion Ltd, Unit 10 Oxford Industrial Park, Mead Road, Yarnton Oxfordshire, OX5 1QU, UK}
\author{E. T.~Gumbrell}
\altaffiliation{Blackett Laboratory, Imperial College London, SW7 2BW, UK}
\altaffiliation{Currently at Lawrence Livermore National Laboratory, CA, USA}
\author{P.~Graham}
\author{S.~Patankar}\altaffiliation{Currently at Lawrence Livermore National Laboratory, CA, USA}
\author{C.~Danson}
\affiliation{AWE, Aldermaston, Reading, West Berkshire RG7 4PR, UK}

\author{C.~Spindloe}
\affiliation{Science and Technology Facilities Council, Rutherford Appleton Laboratory, Harwell Campus, Chilton, Didcot, Oxon, OX11 0QX}

\author{J.~Larour}
\affiliation{LPP, CNRS, Ecole Polytechnique, UPMC Univ Paris 06, Univ. Paris-Sud, Observatoire de Paris, Universit\'e Paris-Saclay, Sorbonne Universit\'es, PSL Research University, 4 place Jussieu, 75252 Paris, France}

\author{M.~Kozlova}
\affiliation{ELI, IOP-CAS, Prague, Czech Republic}

\author{R.~Rodriguez}
\author{J.M.~Gil}
\author{G.~Espinosa}
\affiliation{Universidad de las Palmas de Gran Canaria, Spain}

\author{P.~Velarde}
\affiliation{Instituto de Fusion Nuclear, Universidad Politecnica de Madrid, Spain}


\date{\today}

\begin{abstract}
We present new experiments to study the formation of radiative shocks and the interaction between two counter-propagating radiative shocks. The experiments were performed at the Orion laser facility which was used to drive shocks in xenon inside large aspect ratio gas-cells. The collision between the two shocks and their respective radiative precursors, combined with the formation of inherently 3-dimensional shocks, provides a novel platform particularly suited for benchmarking of numerical codes. The dynamics of the shocks before and after the collision were investigated using point-projection X-ray backlighting while, simultaneously, the electron density in the radiative precursor was measured via optical laser interferometry. Modelling of the experiments using the 2-D radiation hydrodynamic codes NYM/PETRA show a very good agreement with the experimental results.
\end{abstract}


\maketitle




\section{}

Radiative shocks are formed when shocked matter becomes hot enough that radiative energy transfer changes the shock structure. Radiative shocks are ubiquitous in astrophysical phenomena including supernovae \cite{Blondin1998a} and protostellar jets \cite{Hartigan2003}. Photons escaping from the shock can heat and ionize the un-shocked medium ahead of it, leading to the formation of a {\textit{radiative precursor}} \cite{ZeldovichReizerBOOK, DrakeBOOK}. The traditional study of radiative shocks has relied on theory \cite{McClarren2010} and numerical simulations for the interpretation of astrophysical phenomena (e.g. \cite{Innes1987}) and experimental data \cite{Leygnac2006, Gonzalez2009, VanderHolst2012, Fryxell2012a}, which requires the addition of non-local radiative transport to multi-dimensional hydrodynamics. The growth of instabilities and other non-ideal effects can further modify the physics, thus experimental data are essential in order to test these models and improve our understanding of the physics of radiative shocks.

Experiments to produce radiative shocks are typically performed with high-power lasers, which can produce and accelerate plasma flows to velocities $\sim$10-100's km/s (see references in \cite{Drake2011, Michaut2009}). One experimental approach to study such radiative effects consists of producing radiative blast waves by focusing lasers onto a gas-embedded pin \cite{Hansen2006, Edens2010} or into a puffed cluster gas \cite{Edwards2001, Moore2007, Hohenberger2010}. In these cases, the shocks decelerate following a Sedov-Taylor trajectory. An alternative experimental approach consists of focusing lasers onto a foil which, due to laser ablation pressure, acts as a piston that continuously pushes and compresses a static gas inside a tube or gas-cell. In order to maximise radiative effects, the experiments are typically performed in high atomic-number gases such as xenon, at pressures $\lesssim$1 bar \cite{Bozier1986, Bouquet2004, Fleury2002, Reighard2006, Gonzalez2006, Stehle2012}. Results from these piston-driven experiments show the formation of quasi-planar radiative shocks, albeit perturbed by the interaction of the shock with the walls of the tube at velocities $\gtrsim$100 km/s \cite{Doss2009}. These experiments have led to novel applications particularly in the area of laboratory-astrophysics, e.g.~reverse radiative shocks for studies of accretion in cataclysmic variables \cite{Krauland2013a}.

\begin{figure}[t!]
\begin{center}
\includegraphics*[width=8.1cm]{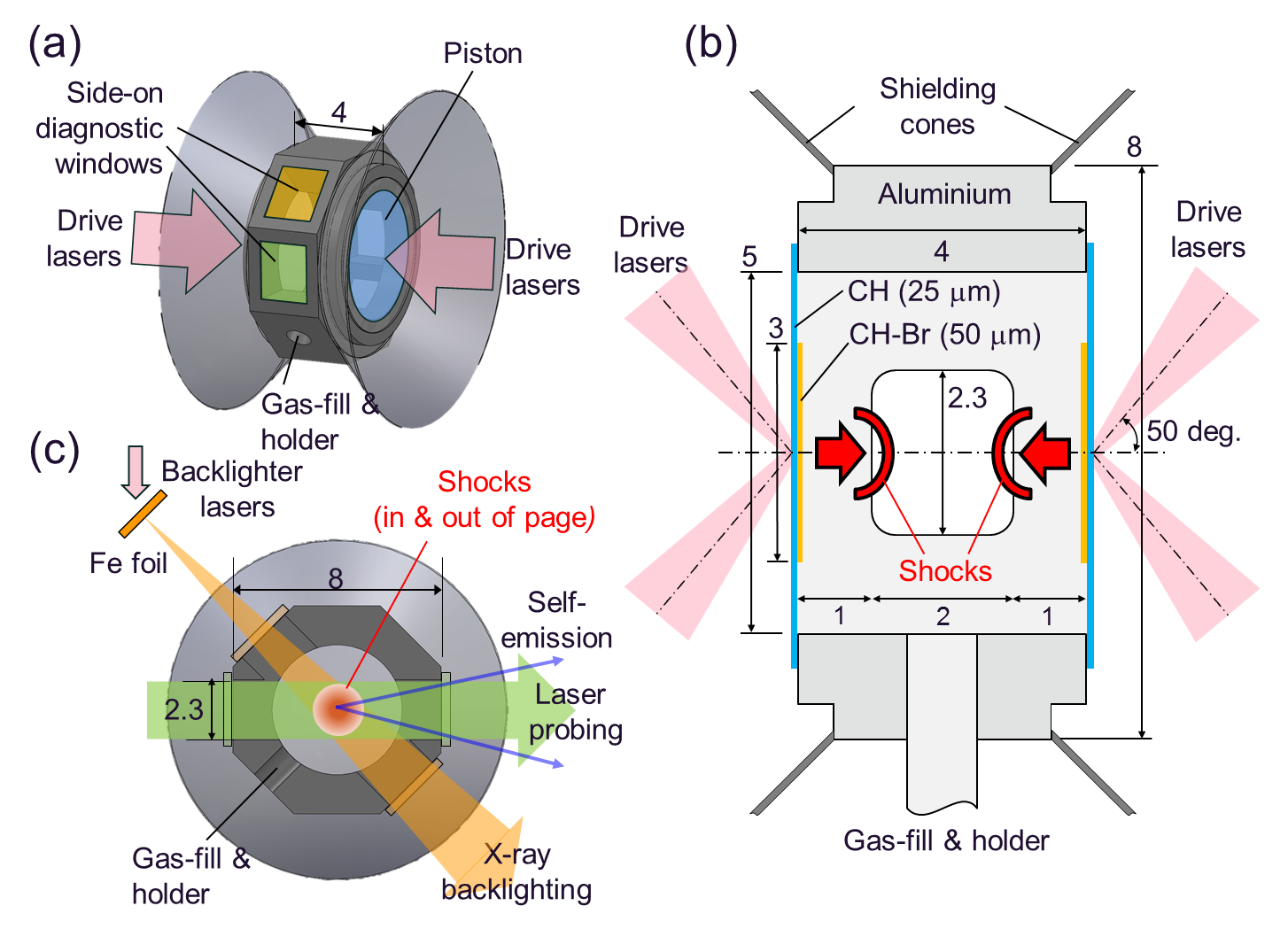}
\caption{Octogonal gas-cells (nominal dimensions in mm). (a) 3-D rendering. (b) Side-on, cut-view. (c) Face-on view and diagnostics.}
\label{fig1}
\end{center}
\end{figure}

In this Letter we report on new experiments designed to investigate the formation of piston-driven radiative shocks. The experiments were designed so the shocks are able to propagate both axially and radially, preventing in this way any interaction of the shocks with the internal walls of the cells and thus leading to a quasi-spherical shock geometry \cite{Koenig2006}. In addition, the collision and interaction between two counter-propagating radiative shocks and their respective radiative precursors is introduced as a radiation-hydrodynamics platform particularly suited for laboratory-astrophysics studies (e.g.~colliding supernova remnants \cite{Velarde2006, Pak2013}) and for numerical benchmarking. 

The experimental setup (Fig.~\ref{fig1}) consisted of octogonal gas-cells with plastic discs attached to opposite ends acting as pistons. The pistons were made up of 25 $\mu$m thick CH ($\rho_{CH}$=0.9~g/cc) with a 50 $\mu$m thick CH-Br ($\rho_{CH-Br}$=1.53~g/cc) attached to the inside surface to prevent early-time pre-heating of the gas from X-rays from the interaction of the drive lasers with the CH. This insured the formation of a radiative precursor comes predominantly from heating of the compressed gas in the shock. 

The experiments were conducted on the Orion laser \cite{Hopps2015}. The shocks were driven using 4 laser beams focused onto each piston simultaneously. Each beam ($\sim$400 J, $\lambda$=351 nm, 1 ns pulse duration) had a flat-topped spatial profile with a $\sim$600 $\mu$m spot diameter and thus a laser intensity of $\sim5\times10^{14}$~W/cm$^2$. Side-on diagnostic access was achieved through two pairs of opposite windows sealed with gas-tight filters suitable for optical and X-ray diagnostics. The gas-cells were filled with xenon to a pressure of P$_0\sim$0.3 bar ($\rho_0\sim$1.6~mg/cc).

The diagnostic setup is presented in Fig.~\ref{fig1}c. One pair of windows was used for point-projection X-ray backlighting (XRBL) imaging of the shocks driven by additional lasers ($\sim$450 J, 500 ps pulse duration) focused onto a 5 $\mu$m thickness iron foil supported on a 20 $\mu$m diameter pinhole that provided spatial resolution \cite{Kuranz2006}. This is comparable to the resolution due to motion blurring ($\sim40~\mu$m) for the XRBL laser pulse duration and a typical shock velocity of v$_s\sim$80 km/s. The resulting emission is dominated by iron He-$\alpha$ transitions (6.7 keV photons)\cite{Phillion1986} which was recorded onto image plates with a magnification of $\sim\times$11.

The second pair of windows were used to perform optical laser interferometry in a Mach-Zehnder configuration with a $\sim$300 mJ, $\lambda$=532 nm, 50 ns pulse duration, $\sim$35 mm beam diameter laser. Two optical streak cameras (100 ns sweep time) recorded interferometry and optical self-emission along the axis of propagation of the shocks. In addition, 4 gated optical intensifiers (GOI) recorded time-resolved, 2-D interferometry images of the shocks at 4 different times per experiment.

\begin{figure}[b!]
\begin{center}
\includegraphics*[width=8.6cm]{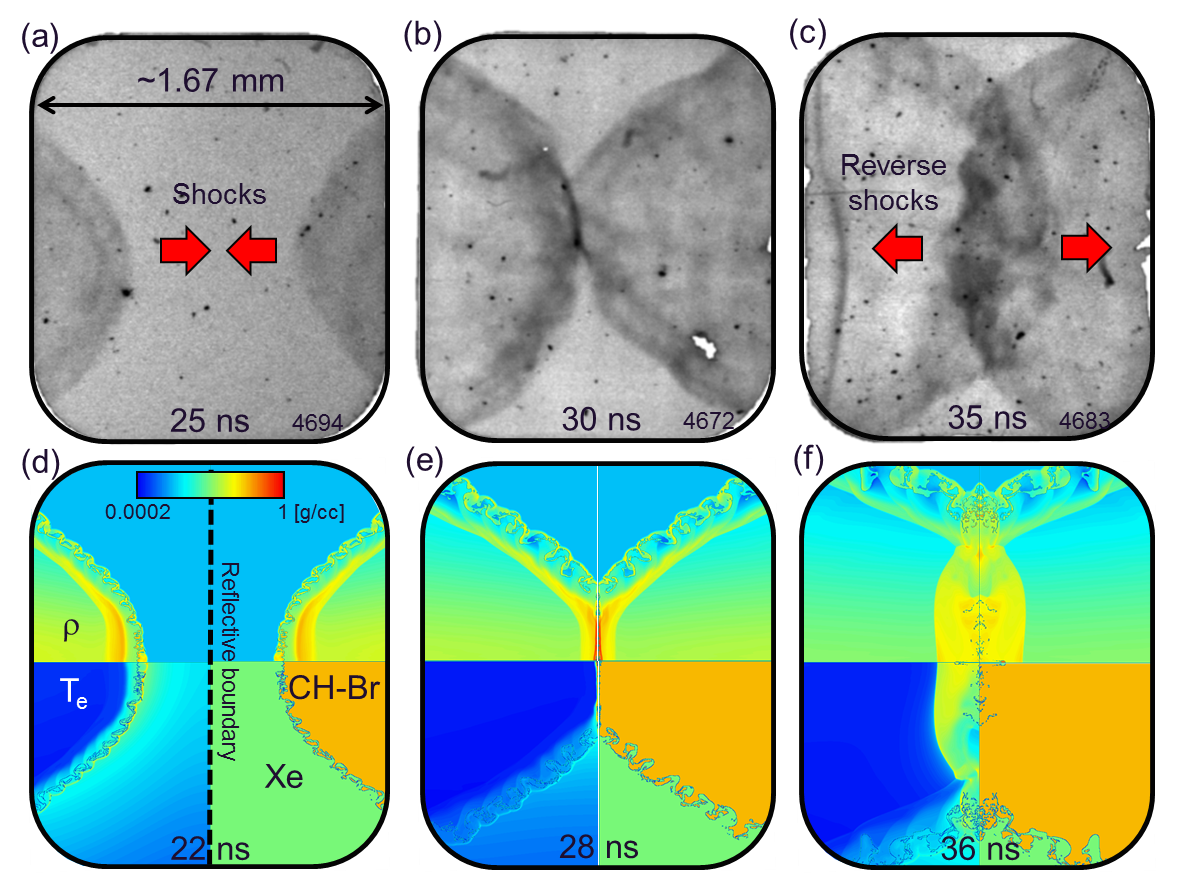}
\caption{Counter-propagating shock dynamics at different times from: (a)$-$(c) experimental X-ray backlighting, (d)$-$(f) 2-D numerical simulations. Each simulation image shows mass density (top-half, log scale), electron temperature (bottom-left quadrant, linear scale) and materials (bottom-right quadrant). The colorbar used to represent mass density in (d) also displays linear values of electron temperature with ranges: (d) 0$-$35 eV, (e) 0$-$60 eV, (f) 0$-$40 eV.}
\label{fig2}
\end{center}
\end{figure}

Figs.~\ref{fig2}a-c show results from XRBL at 25, 30 and 35 ns. The shocks are seen as round-shaped features coming into the field of view of the windows from each side, with darker tones representing stronger X-ray absorption (i.e. higher mass density). Their head-on collision is seen at 30 ns followed by the formation of reverse shocks as dense structures at the center of the window at 35 ns. These results indicate a shock velocity of $v_s\sim75\pm25$ km/s (i.e.~a shock displacement of $\sim0.25-0.5$ mm in 5 ns) and a reverse shock velocity of the order of $v_{rs}\sim$30 km/s.

\begin{figure}[t!]
\begin{center}
\includegraphics*[width=8.7cm]{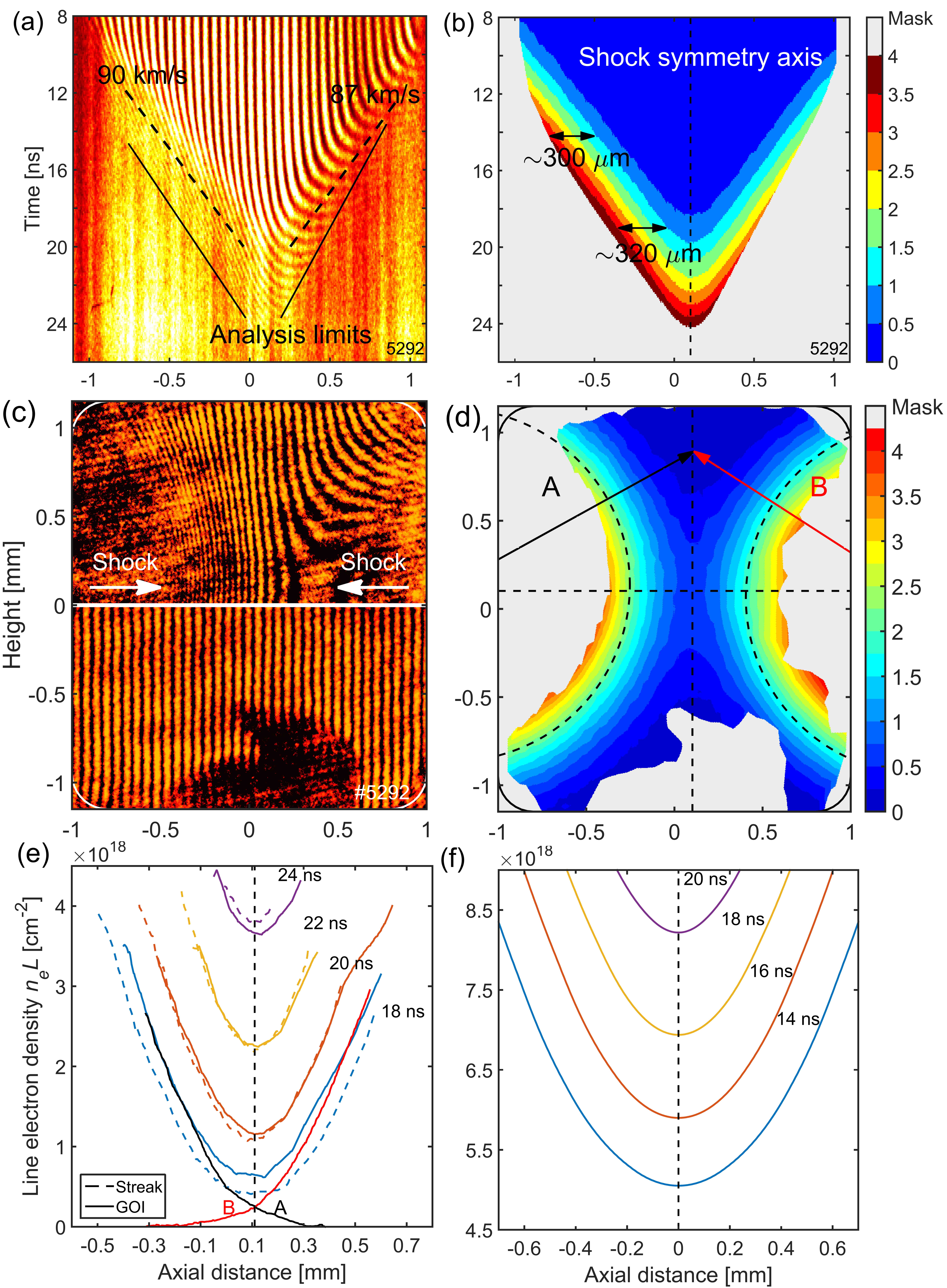}
\caption{Electron density measurements in the radiative precursors. (a)-(b) 1-D axial streak interferometry results, respectively (a) raw data and (b) analysis of (a) resulting in line electron density $n_eL$ ($\times10^{18}$~cm$^{-2}$) as a function of time. The dashed lines on (a) mark values of $n_eL\sim1.5\times10^{18}$~cm$^{-2}$. (c)-(d) 2-D GOI results at 18 ns. The top-half of (c) shows raw data and the bottom-half a pre-shot interferogram. (d) 2-D $n_eL$ map from analysis of (c). (e) Axial profiles of $n_eL$ at 18, 20, 22 and 24 ns from (b) and (d). Also shown are $n_eL$ profiles at 18 ns off-axis (positions marked as \textit{A} and \textit{B} in (d)). (f) Simulated axial profiles of $n_eL$ at 14, 16, 18 and 20 ns.}
\label{fig3}
\end{center}
\end{figure}

Figs.~\ref{fig2}d-f show 2-D axisymmetric simulations with the radiation-hydrodynamics codes NYM/PETRA using the same initial experimental conditions as in the experiments (2 $\mu$m resolution). NYM \cite{Roberts1980a} is a Lagrangian code with multi-group implicit Monte-Carlo X-ray transport and full laser-interaction physics used to model the laser-piston interaction. These simulations were linked and mapped to the Eulerian code PETRA (typically after 5 ns) \cite{Youngs1984}, using multi-group X-ray diffusion to study the late-time plasma behaviour. The opacities and equations of state for the multi-material piston and xenon were taken from SESAME tables. The counter-propagating shock collision was simulated using a fully reflective boundary at the centre of the diagnostic window (shown schematically in Fig.~\ref{fig2}d) for the plasma flow and radiation. 

The simulations accurately reproduce the overall shock dynamics with an uncertainty of up to $\sim$3 ns, which can be attributed to shot-to-shot experimental variations in the targets and laser energy. Simulated mass density reproduces the increase in density seen at the shock front in XRBL (see Figs.~\ref{fig2}a,d), which can be attributed to regions of post-shock xenon followed by CH-Br. The plots of materials indicate the typical width of the post-shock xenon is $\sim40~\mu$m, i.e.~in the limit of the diagnostic resolution due to motion blurring. The simulations show the shock front as an unstable, rippled layer due to the growth of hydrodynamic instabilities mediated by strong radiative cooling in the shock which lead to an increase in its density, thus making the interface with the upstream, un-shocked xenon unstable. Simulated electron density in Fig.~\ref{fig2}d at 22 ns shows significant heating ahead of the shock due to the formation of the radiative precursor.

\begin{figure*}[t!]
\begin{center}
\includegraphics*[height=5.5cm]{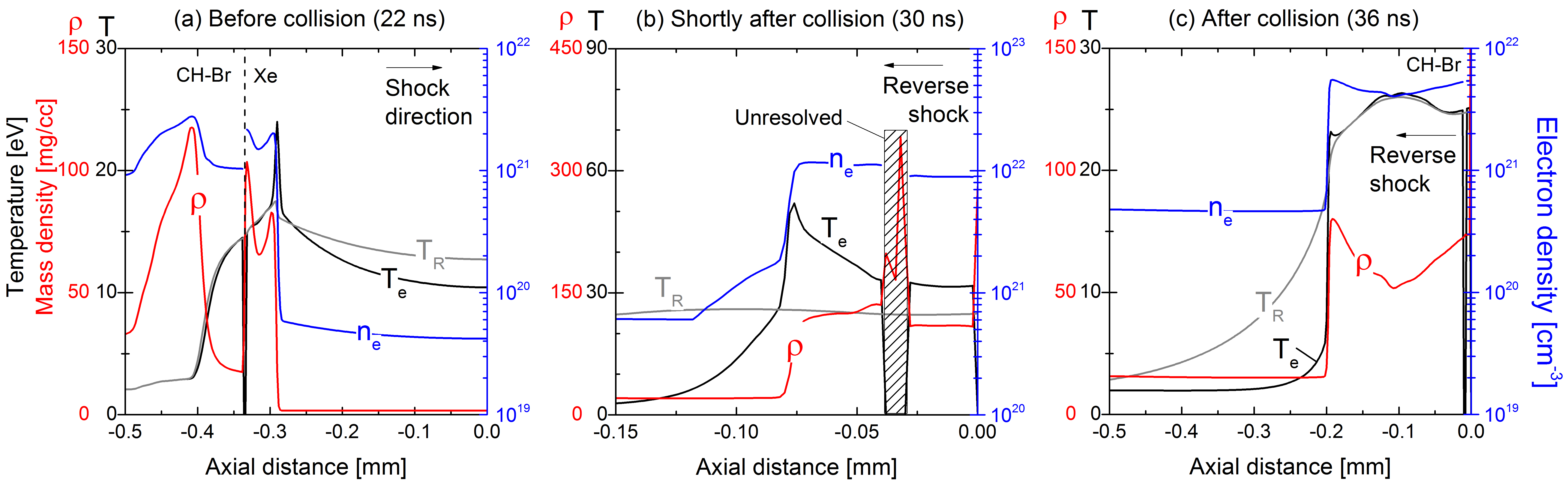}
\caption{Axial profiles of mass density ($\rho$), electron density ($n_e$), electron temperature ($T_e$) and radiation temperature ($T_R$) from 2-D simulations of counter-propagating radiative shocks at: (a) 22 ns (before the collision, see Fig.~\ref{fig3}d), (b) 30 ns (shortly after the collision, note change in X and Y scales), (c) 36 ns (after the collision, see Fig.~\ref{fig3}f). The axial distance is taken from the centre of the diagnostic window, which marks the position of a reflective boundary used for the simulations.}
\label{fig4}
\end{center}
\end{figure*}

As the XRBL diagnostic is sensitive to variations in mass density, it does not provide information on the radiative precursor which is characterized by changes in temperature and ionization. Thus the radiative precursor was studied by measuring the electron density ahead of the shock with laser interferometry. Fig.~\ref{fig3} shows results from 1-D axial streak imaging (Figs.~\ref{fig3}a-b) and 2-D time-resolved GOI imaging (Figs.~\ref{fig3}c-d). The displacement of the interference fringes from their initial undisturbed position (straight lines in Figs.~\ref{fig3}a,c) is proportional to the line electron density $n_eL$, i.e.~the electron density integrated along the length of the plasma being probed. As time progresses, the fringe contrast decreases and the displacement cannot be traced accurately as the laser goes through regions near the shocks, characterized by strong absorption and large spatial gradients of electron density. The resulting $n_eL$ maps (Figs.~\ref{fig3}b,d) were obtained using the technique described in \cite{Swadling2013a}. The streak interferometry results in Fig.~\ref{fig3}b show that isocontours between $n_eL=1\times10^{18}$~cm$^{-2}$ and $3.5\times10^{18}$~cm$^{-2}$ maintain an approximately constant separation from $\sim12-20$~ns indicating that, during these times, the radiative precursor reaches a steady state with a characteristic extent of $\sim300-350$~$\mu$m. Such quasi-stationary radiative shocks have only been previously observed in 1-D experiments \cite{Stehle2010} and in 2-D numerical simulations\cite{Michaut2009}. By following a fixed value of $n_eL=$1.5$\times10^{18}$~cm$^{-2}$ a characteristic precursor ``velocity'' of $\sim90$~km/s is estimated, in agreement with the shock velocity estimated from XRBL results.

Results from 2-D $n_eL$ at 18 ns in Figs.~\ref{fig3}c-d show similar features as those seen in XRBL results in Fig.~\ref{fig2}a, however with this diagnostic the shocks cannot be accurately resolved and are seen as diffuse regions. The analysis in Fig.~\ref{fig3}d shows that the isocontours of $n_eL$ between $1-3\times10^{18}$~cm$^{-2}$ can be well approximated as concentric circles (e.g.~$n_eL=2\times10^{18}$~cm$^{-2}$ shown as dashed circles) indicating that the precursors have a spherical shape at this particular time. This allows defining origins for radial axes of symmetry for both shocks, roughly aligned with the horizontal shock symmetry axis (shown as straight dashed lines). The radial symmetry allows extracting profiles of $n_eL$ towards regions off-axis, where little interaction with the counter-propagating precursor is expected. Under this approximation, it is possible to extract the expected $n_eL$ profiles for a single-drive shock (labelled as \textit{A} and \textit{B} in Fig.~\ref{fig3}d). 

Fig.~\ref{fig3}e shows a comparison of axial profiles of $n_eL$ from streak and GOI imaging at 18, 20, 22 and 24 ns, showing a very good agreement between both diagnostics. Discrepancies are probably due to uncertainties in defining a ``zero'' level of $n_eL$ \cite{Swadling2013a} in the analysis of GOI data. Comparison between the profiles on- and off-axis at 18 ns in Fig.~\ref{fig3}e show the radiative precursors have very similar values away from the vertical symmetry axis and, as they reach the collision in the centre, it leads to an effective increase in $n_eL$. This increase is consistent with the values obtained by doing the sum between the two off-axis profiles \textit{A} and \textit{B}. 

The axial spatial distribution of $n_eL$ from the two radiative precursors in Fig.~\ref{fig3}e can be compared to results from 2-D simulations shown in Fig.~\ref{fig3}f. The simulations overall match the spatial distribution of $n_eL$ in the experiments with a time difference of 4 ns (14$-$20 ns in simulations, 18$-$24 ns in the experiments). Moreover, simulations overestimate the experimental values of $n_eL$ by a constant value of $\sim4.5\times10^{18}$ cm$^{-3}$, which is consistent with experiments having a 3-D distribution of electron density instead of 2-D in simulations, thus lower values should be expected in reality as discussed in \cite{Vinci2006b}.


In order to get a better understanding of the collision between the two counter-propagating radiative shocks, axial profiles from 2-D numerical simulations in Figs.~\ref{fig4}a-c show, respectively, the plasma conditions at 22 ns (pre-collision), at 30 ns (shortly after the collision), and at 36 ns (post-collision). In Fig.~\ref{fig4}a, the dip in electron temperature at $\sim$-0.34 mm from the reflective boundary (at 0 mm) marks regions of CH-Br and Xe, whereas the peak in electron temperature of $T_e\sim24$~eV at $\sim$-0.29 mm marks the position of the shock front. The radiative precursor is seen ahead of the shock with a peak temperature of $T_e\sim16$ eV, decreasing to $T_e\sim10$~eV on the axis of the window. Similar electron temperatures pre- and post-shock indicate the shock is supercritical \cite{ZeldovichReizerBOOK}. The post-shock xenon temperature $T_{ps}$ agrees with estimates presented in \cite{Drake2011} done by balancing the fluxes of radiation and kinetic energy of the incoming flow ($2\sigma T_{ps}^4=\rho_0 v_s^3/2$). Here $\rho_0=1.6$~mg/cc is the initial Xe density, $v_s\sim80$ km/s is the measured shock velocity and $\sigma=5.67\times10^{-8}$~Wm$^{-2}$K$^{-4}$ is the Stefan-Boltzmann constant, resulting in $T_{ps}\sim20$~eV. 

The simulated mass density in the post-shock xenon region in Fig.~\ref{fig4}a at 22 ns shows a double peak which reflects the spatial variations in density due to the formation of hydrodynamic instabilities seen in 2-D images (ripples). Thus a lower-boundary for the post-shock compression can be estimated by taking the density at the through (which remains constant between 22$-$26 ns) of $\rho_{ps}\sim60$~mg/cc, resulting in a compression of $\rho_{ps}/\rho_{0}\sim38$. On a first approximation, the post-shock compression can also be estimated experimentally from XRBL results in Fig.~\ref{fig2}a by taking the ratio of absorbed X-ray intensity at a point through the post-shock $I_{ps}$ respect to the intensity through the undisturbed ambient xenon $I_0$ via the expression $\rho_{ps}/\rho_{0}=1+[ln(I_{0}/I_{ps})/(\sigma_{Xe}L_{Xe}\rho_0)]$, where $\sigma_{Xe}=505$ cm$^2$/g is the mass attenuation coefficient for Xe at 6.7 keV and $L_{Xe}\sim0.3$ mm is the transverse length of the xenon post shock. Typical values of $I_{0}/I_{ps}$ are $\sim1.1$, and thus from different shots $\rho_{ps}/\rho_{0}\sim6\pm2$. It should be noted that this estimate is heavily constrained by the resolution of the XRBL diagnostic (of the order of the extent of the post-shock region) and the possible emission of hard X-rays ($>$10 keV) from the backlighter \cite{Krauland2012} which could affect the intensity measurements. Overall the compressions that characterise the post-shock are higher than the ideal, non-radiative compression of 4$\times$ \cite{Bouquet2004}, which indicates that radiative losses play a significant role in the shock dynamics.

Results from simulations post-collision, e.g.~from 28 ns onwards (Figs.~\ref{fig4}b,c and Figs.~\ref{fig2}e,f) indicate the reverse shock is formed mostly of piston material (CH-Br) with an almost unresolved xenon region close to the reflective boundary. The post-shock density in the reverse shock $\rho_{prs}$ can be estimated by using the Rankine-Hugoniot relations for a reverse shock \cite{Fortmann2012, Pak2013} as $\rho_{prs}=((v_{rs}+v_{ps})/v_{rs})\rho_{ps}$, where $v_{rs}$ is the reverse shock velocity in the laboratory frame (measured as $v_{rs}\sim30$ km/s), $v_{ps}$ is the post-shock velocity in the laboratory frame given by $v_{ps}=v_s(\rho_{ps}-\rho_0)/\rho_{ps}\approx73$ km/s for a post-shock mass density from simulations of $\rho_{ps}\sim60$ mg/cc, resulting in $\rho_{prs}\sim206$ mg/cc. This estimate is in line with simulation results at 28$-$30 ns (e.g.~Fig.~\ref{fig4}b), which indicate $\rho_{prs}\sim110-190$ mg/cc, with the caveat that strong mixing between Xe and CH-Br should be predominant at these times. Simulations at 36 ns in Fig.~\ref{fig4}c show the reverse shock as an extended region with a half-width of $\sim0.2$ mm and with an approximately constant mass density and temperature of $\rho\sim50$ mg/cc and $T_e\sim$25 eV respectively. This region drives a strong reverse-radiative precursor evidenced by an increased radiation temperature $T_R$.

In summary, we presented a new study of laser-piston driven radiative shocks in xenon characterised by simultaneous experimental measurements of the dynamics of the shock region and the radiative precursor. This experimental set-up allows studying the collision between two counter-propagating radiative shocks as a radiation-hydrodynamics platform to study complex physics, particularly suited for numerical benchmarking. Simulations are able to accurately reproduce the experimental results, and we hope the first results in this Letter can be used as a test bed for other codes (e.g.~3-D radiative codes) to investigate, for instance, the effect of the interaction between the precursors and the formation of the reverse shocks. Similarly, we intend for future experiments to corroborate the estimates presented here by measuring the precursor and post-shock plasma conditions in xenon with, e.g. X-ray Thomson scattering, which so far has only been done for radiative shocks in argon \cite{Visco2012}. 

\begin{acknowledgments}

This work was supported by STFC and AWE through their academic access programme, in part by The Royal Society, EPSRC and Labex PLAS@PAR. The authors would like to acknowledge Robert Charles, Jim Firth, Paul Treadwell, Rob Johnson, David Hillier, Nick Hopps and the entire Orion team for their help and support during the experiments.

\end{acknowledgments}


%

\end{document}